\documentclass[useAMS,usenatbib]{mn2e}
\usepackage{amsmath, amssymb}
\usepackage{graphicx}
\usepackage{xcolor}
\usepackage{textcomp,gensymb}
\numberwithin{equation}{section}

\newcommand{\Mpc}{\ensuremath{\mathrm{Mpc}}}
\newcommand{\eV}{\ensuremath{\mathrm{EeV}}}

\renewcommand{\ga}{\ensuremath{\gamma}}

\newcommand{\msun}{\ensuremath{M_{\odot}}}

\newcommand{\ten}[1]{\ensuremath{10^{#1}}}

\begin{document}
\bibliographystyle{mn2e}

\title[TDEs and UHECR Hot Spots]{Ultra-high-energy-cosmic-ray
     hotspots from tidal disruption events}
\author[Daniel N. Pfeffer, Ely D. Kovetz, and Marc
     Kamionkowski]{Daniel N. Pfeffer, Ely D. Kovetz and Marc
     Kamionkowski \\
     Department of Physics and Astronomy, Johns Hopkins
     University, Baltimore, MD 21218 USA}

\maketitle
\label{firstpage}

\begin{abstract}
We consider the possibility that tidal disruption events (TDEs) caused by supermassive
black holes (SMBHs) in nearby galaxies can account for the ultra-high-energy
cosmic-ray (UHECR) hotspot reported recently by the Telescope Array (TA) and the
warm spot by Pierre Auger Observatory (PAO). We describe the expected cosmic-ray
signal from a TDE and derive the constraints set by the timescale for dispersion due
to intergalactic magnetic fields and the accretion time of the SMBH. We find that
TDEs in M82 can explain the hotspot detected by the TA regardless of whether the
UHECRs are composed of protons or heavier nuclei. We then check for consistency of
the hot and warm spots from M82 and Cen A with the full-sky isotropic signal from
all SMBHs within the GZK radius. This analysis applies to any scenario in which the
hot/warm spots are real and due to M82 and Cen A, regardless of whether TDEs
are the source of UHECRs. We find that the isotropic flux implied by the luminosity
density inferred from M82 and Cen A is bigger than that observed by roughly an order
of magnitude, but we provide several possible explanations, including the possibility
of a local overdensity and the possibility of intermediate-mass nuclei in UHECRs, to
resolve the tension.
\end{abstract}

\section{Introduction}

In the past decade the ability to observe ultra-high-energy
cosmic rays\footnotemark (UHECRs) has increased significantly
with the advent of the Pierre Auger Observatory (PAO) and the
Telescope Array (TA).  Recently, both the TA and the PAO have
detected regions of excess UHECRs as compared to an isotropic
background  \citep{Abbasi:2014lda,PierreAuger:2014yba}, with
statistical significances of $\gtrsim 3 \sigma$ and $\gtrsim 2
\sigma$, respectively. 

\footnotetext{For the purpose of this paper, UHECRs will be
defined as cosmic rays with energies above 57 \eV.}

The sources of UHECRs are still unknown.  One possibility is
active-galactic-nucleus jets \citep{Abraham:2007bb}.  However,
\citet{Farrar:2008ex} derived a relation between the AGN
electromagnetic luminosity and its UHECR luminosity.
\cite{Zaw:2008is} then used the Veron-Cetty and Veron (VCV)
catalogue, along with this luminosity relation, to infer that
the observed AGN are not luminous enough to explain the full-sky
UHECR flux.  Gamma-ray bursts (GRBs) are also capable of
producing UHECRs \citep{Waxman:1995vg}, but they would have to
have a rather flat spectrum of UHECRs produced by an individual
GRB and would have to yield over 100 times more energy to UHECRs
than to photons in order to explain the full-sky flux
\citep{Farrar:2008ex}.

We consider a third mechanism as the dominant source of UHECRs,
namely tidal disruption events (TDEs).  A star is disrupted by a
super massive black hole (SMBH) when it passes by close	enough
that tidal forces overcome the binding energy of the star.
Some fraction of the star then becomes bound to the SMBH
and forms a short-lived accretion disk which produces an intense
flare\footnotemark \citep{Rees1988}.  Some of the TDEs produce jets
which were first proposed as a source of UHECRs in
\citet{Farrar:2008ex}, and then expanded upon in
\citet{Farrar:2014yla}, which showed that they can generate the
luminosity required to account for the full-sky UHECR flux.

\footnotetext{The SMBH does not need to be an
AGN---i.\ e.\ actively accreting from the accretion disk---in
order for the disruption to cause rapid accretion.  The
in-falling gas from the disrupted star could form an accretion
disk with rapid  accretion resulting in a relativistic jet
outflow \citep{Farrar:2008ex}.}

In 2014, the TA reported a ``hot spot'' of UHECRs
\citep{Abbasi:2014lda} in a circle of radius 20\degree{}, centred
at a right ascension of 146.\degree{}7 and declination of
43.\degree{}2.  \citet{He:2014mqa} tried to
identify possible extragalactic sources for the hot spot, taking
into account possible deflection of the UHECRs by Galactic and
intergalactic magnetic fields.  After accounting for random
deflections by stochastic intergalactic magnetic fields (IGMFs),
they drew a straight line through the images of the different
rigidity bins of the events in the hotspot, expecting the source
to lie along this line. Two possible sources were  identified,
M82 and Mrk 180. While Mrk 180 is located roughly 185 \Mpc{}
away, near the GZK radius, and is thus unlikely to be the
source, M82 is a starburst galaxy only 3.8 \Mpc{} away
\citep{Karachentsev:2005gp} and moreover has a \(\sim
3\times\ten{7}\, \msun\) SMBH at its centre
\citep{Gaffney:1993zz}.  The SMBH does not exhibit any AGN
activity.

Likewise, the Pierre Auger Observatory has noted a ``warm
spot,'' an excess of events in the direction of Centaurus A (Cen
A).  Cen A is also (coincidentally) approximately 3.8 \Mpc{}
away \citep{Harris:2010we}, with a SMBH with a mass estimated to be
\(5 \times \ten{7} \msun\).  Unlike M82's this SMBH does exhibit
AGN activity and has a contentious jet.

In this paper we investigate whether the TA hotspot can be
explained by TDEs in M82.  We first derive basic constraints to
the model parameters from timescale and energetic arguments.  We
surmise that the UHECR hot spot is in roughly steady state in
which the UHECR flux results from several TDEs that have
occurred within the timescale for dispersion of a burst signal
due to deflections in the Galactic and intergalactic magnetic
fields (although we do briefly consider the possibility that the
hot spot arises from a single burst.)  The model parameters are
then a TDE rate in M82 and an efficiency for conversion of the
stellar rest-mass energy into UHECRs.  We find find that there
are indeed TDE rates, with plausible efficiencies, that are
large enough to account for the hot spot yet small enough so
that the time-averaged accretion rate onto the SMBH is still
sub-Eddington.  This conclusion follows if UHECRs are composed
of protons and if there are also heavier nuclei in UHECRs,
although the consistent parameter space is a bit
smaller for heavier nuclei.  Similar arguments apply to the warm
spot from Cen A.  We then investigate whether the UHECR
luminosity density implied by the observed fluxes from the
SMBHs in M82 and Cen A is consistent with the isotropic UHECR
that is observed.  We find that the isotropic flux inferred in
this way is higher, by about a factor of 16, than the observed
isotropic flux, but we point out several factors that might
alleviate the apparent discrepancy.

The rest of this paper is organized as follows. In Section
\ref{sec:data}, we review briefly the evidence of the TA hot
spot and the PAO warm spot and provide the fiducial values we
use for the hot-spot and warm-spot fluxes as well as the
isotropic UHECR intensity.  In Section \ref{sec:timescales} we
discuss the constraints to TDE scenarios for the UHECRs hot/warm
spots that arise from energetics and timscale considerations.
In Section \ref{sec:isotropic} we consider constraints to the
scenario that arise from consistency of the hot/warm-spot fluxes
with the isotropic UHECR intensity.  In Section
\ref{sec:discuss} we summarize, review the successes and
weaknesses of the TDE explanation for the hot/warm spots,
consider some possible future measurements, and close with some
speculations.

\section{The hot and warm spots} 
\label{sec:data}

The TA Collaboration reports evidence \citep{Abbasi:2014lda} for
a UHECR excess in a circle of  $20^\circ$ radius.  \cite{Fang:2014uja}
estimate the specific (number) intensity $J_H$ in this hot spot to be,
\begin{equation}
     E^2 J_H = (4.4\pm 1.0)\times 10^{-8} \, {\rm GeV}\,{\rm
     cm}^{-2}\, {\rm s}^{-1}\, {\rm sr}^{-1},
\end{equation}
at an energy $E=10^{19.5}$ eV.  The hot-spot energy flux in
UHECRs with energies $>57$ EeV is $F_{\rm hs} =
\Omega_{20^\circ}\,\int_{57 \, {\rm EeV}}^\infty \, E\, J(E)\,
dE$, where $\Omega_{20^\circ}\simeq 0.38$ sr is 
the hot-spot solid angle.  The energy dependence of $J(E)$ at
energies above 57 EeV is, however, is quite uncertain in the hot
spot, and even for the full-sky flux [see, e.g., Fig.~7 in
\citet{Kistler:2013my}, which shows considerable disagreement
between PAO and TA at the highest energies].  We therefore take
the energy flux in the hot spot to be,
\begin{equation} 
\label{eq:hotSpotObs}
     F_{\mathrm{hs}} = 1.7\times 10^{-8}\, F_{1.7} ~{\rm
     GeV}~{\rm cm}^2~{\rm s}^{-1},
\end{equation}
and keep the quantity $F_{1.7}$, which parametrizes our
uncertainty in the flux, in our expressions below.

Likewise, we take the observed isotropic (energy) intensity
above 57 EeV to be $I_{o} = 7.9 \times \ten{-9}$ GeV
cm\(^{-2}\) s\(^{-1}\) sr\(^{-1}\) \citep{Kistler:2013my}.
Again, to be consistent with our treatment of the hot-spot flux,
we take this to be the value of $E^2\,J(E)$ at
$E=10^{19.5}$~eV.  This isotropic flux appears below only in
comparison to the hot-spot flux, and so it is appropriate to
treat the full-sky flux in same way as the hot-spot flux.

We estimate the UHECR energy flux from Cen A implied by the PAO
warm spot as follows:  \citet{Abreu:2010ab} finds 13 events
within a circle of radius $18^\circ$, where 3.2 are expected
from an isotropic distribution.  We thus take the energy flux from Cen
A to be $(13-3.2)/3.2 \approx 3$ times the isotropic energy flux
in that circle, or
\begin{equation} 
\label{eq:CenAObs}
     F_{\mathrm{ws}} = 7.6\times 10^{-9}\, {\rm
     GeV}~{\rm cm}^{-2}2~{\rm s}^{-1},
\end{equation}
keeping in mind the considerable uncertainty in this value.

\section{Time scales and energetics}
\label{sec:timescales}

Our aim here is to understand whether TDEs from accretion of
stars onto the SMBH in M82 may be responsible for the UHECR hot
spot.  We begin with some basic considerations, starting with time
scales.

The hot spot is observed to be spread over an angular region of
size $\theta\sim 20^\circ$.  Such a spread is to be expected due
to scattering in turbulent intergalactic magnetic fields (IGMFs)
as the UHECRs propagate the 3.8~Mpc distance from M82, and there
may be additional scattering (particularly for iron nuclei) from
magnetic fields in the Milky Way.  The rms deflection angle
for a UHECR of charge $Z$ is \citep{He:2014mqa},
\begin{equation} 
\label{eq:magDisp}
     \delta_{\rm{rms}} \approx 3.6\degree\, Z E_{20}^{-1}
     r_{100}^{1/2} \lambda_{\Mpc}^{1/2} B_{\rm{nG}, \rm{rms}},
\end{equation}
where \(B_{\rm{nG}, \rm{rms}}\) is the rms strength of the
magnetic field in nG, $E_{20}$ is the UHECR energy in units of
$10^{20}$ eV, \(r_{100} = r/100\) Mpc is the distance over
which the magnetic fields act, and $\lambda_{\Mpc}$ is the
magnetic-field coherence length in units of Mpc.  Consider first
scattering in Galactic magnetic fields.  Characteristic values
might then be $\lambda_{\Mpc} \sim 10^{-3}$,
$r_{100}\sim10^{-4}$, and $B_{\rm nG}\sim 10^3$, implying
Galactic deflection angles $\delta_{\rm rms} \sim 1^\circ Z$, in
rough agreement with \citet{Giacinti:2010dk} and
\citet{Farrar:2012gm}.  We thus infer that for iron nuclei all
the scattering could conceivably arise from Galactic magnetic
fields, although it is more likely that if the UHECRs are
protons, scattering in IGMFs is more important.

Either way, scattering in magnetic fields also gives rise to a
spread \citep{Waxman:1995vg,Farrar:2014yla}
\begin{eqnarray}
     \tau &\simeq& 3\times 10^5 \left( \frac{r_{100} B_{\rm
     nG}}{E_{20}} \right)^2 \lambda_{\Mpc} Z^2\, {\rm yr}
     \nonumber \\
     & \simeq& 3.5\times 10^5\, \left(\frac{\delta_{\rm
     rms}}{3.6^\circ} \right)^2 r_{100}\, {\rm yr}.
\end{eqnarray}
Thus, if all the scattering takes place in the Milky Way, for
which $r_{100}\sim 10^{-4}$,
then $\delta_{\rm rms}\sim20^\circ$ implies a dispersion of
$\tau \sim 1000$ yr in the UHECR arrival times.  If scattering occurs
primarily in IGMFs, then the spread in arrival times is more
like $\tau \sim 4\times 10^5$.  {\it We thus infer that UHECRs
are spread in arrival time by some magnetic-dispersion
timescale $1000\,{\rm yr} \lesssim \tau \lesssim 4\times
10^5\,{\rm yr}$, with protons more likely to fall near the higher
end and iron nuclei closer to the lower end.}

We now consider energetics.  If the observed flux in $E>57$~EeV
UHECRs in the hot spot is $F_{\rm hs} \simeq 1.7\times
10^{-8}\, F_{1.7}$~GeV~cm$^{-2}$~sec$^{-1}$, then the implied
isotropic-equivalent source luminosity is $L=4 \pi D^2 F \simeq
8.3 \times 10^{-7}\,F_{1.7}\,M_\odot c^2$~yr$^{-1}$ (where
$D=3.8$ Mpc is the distance).  If the
observed UHECRs are due to a single TDE spread out over a time
$\tau$, then the isotropic-equivalent energy implied with
$\tau \simeq 1000$~yr (more likely for iron nuclei) is $8.3\times 
10^{-4}\,F_{1.7}\, M_\odot c^2$.  If the dispersion time 
is $\tau\simeq 4\times 10^5$~yr, the more likely value for
protons, then the isotropic-equivalent energy is
$0.33\,F_{1.7}\,M_\odot c^2$.  Of course,
if the TDE is beamed into a solid angle that subtends a
fraction $\Omega_{\rm jet} \sim 0.1$ of $4\pi$, then the energy
requirements can be relaxed by a factor $\sim 10$.  Still, we
conclude that if UHECRs are iron nuclei, the hot spot is
conceivably due to a single burst.
If the UHECRs are protons, the energetics
are prohibitive, unless the Milky Way magnetic-field
parameters are altered so that the angular spread in the hot spot
arises from scattering in the Milky Way.  Even if the energetics
can somehow be worked out, the notion that we are seeing a hot
spot just from M82 because of some chance occurrence (an
extroardinarily energetic TDE at just the right time) is
unsatisfying, and even more unsatisfying if we must also explain
the warm spot as some similar chance occurrence in Cen A.

Another possibility is that the observed hot spot arises not
from a single TDE, but from a number of TDEs in M82.  This may
occur if the dispersion $\tau$ in arrival times exceeds the
typical time  $\Delta t$ between TDEs in M82.  If so, then we
are seeing UHECRs from $N\simeq (\tau/\Delta t) \gtrsim 1$
bursts at any given time.  The hot-spot flux in this case will
vary by a fractional amount $\sim N^{-1/2}$ over timescales
$\sim\tau$.  However, over the $\sim 5$-yr observation, the
observed flux will remain effectively constant.  This scenario,
as we will now show, is plausible.

We suppose that stars (which we assume for simplicity to all have
a mass $M_\odot$) are swallowed by the SMBH with a rate
$\Gamma$.  We then suppose that only a fraction $\zeta$
produce the type of jets that can accelerate UHECRs and that a
fraction $\xi$ of the stellar rest-mass energy $M_\odot c^2$
goes into UHECRs.  We further suppose that the UHECR emission
may be beamed into a fraction $\Omega_{\rm jet}$ of the
$4\pi$ solid angle of the sphere.  In order to obtain the
observed UHECR hot-spot flux in steady state, we require that
stars be swallowed by the SMBH at a rate,
\begin{equation}
     \Gamma = 8.3\times 10^{-7}\, \left( \frac{\Omega_{\rm jet}
     F_{1.7}}{\xi \zeta} \right)\, {\rm yr}^{-1}.
\label{eqn:swallowingrate}
\end{equation}
The mean time between UHECR-producing events is 
\begin{equation}
     \Delta t = (\zeta \Gamma)^{-1} = 1.26\times 10^6\,
     \frac{\xi}{\Omega_{\rm jet} F_{1.7}} \, {\rm yr}.
\end{equation}
If this mean time is to be smaller than the magnetic-dispersion
time $\tau$, we require
\begin{equation}
     \frac{\xi}{\Omega_{\rm jet} F_{1.7}} \lesssim 0.31\,
     \tau_4,
\label{eqn:dtr}
\end{equation}
where $\tau_4$ is the magnetic-dispersion time in units of
$4\times 10^5$ yr.

We now compare the mass-accretion rate implied by
equation~(\ref{eqn:swallowingrate}) with the Eddington rate
$\dot M = L_{\rm Edd}/c^2 \simeq 3.8\times 10^{45}\,M_3\,{\rm
erg}\,{\rm s}^{-1}/c^2$, where $M_3$ is the SMBH mass in units
of $3\times 10^7\, M_\odot$, for M82.  Assuming that half of the
disrupted star's mass is accreted, we find that the
mass-accretion rate is smaller than Eddington if
\begin{equation}
     \frac{\xi}{\Omega_{\rm jet} F_{1.7}} \gtrsim 6.0 \times
     10^{-6} M_3^{-1} \zeta^{-1}.
\label{eqn:Eddrequire}
\end{equation}
It is not, strictly speaking, required that this condition be
respected.   It is conceivable that a SMBH could appear
quiescent, even with a super-Eddington time-averaged
mass-accretion rate, if the accretion is episodic.  Still, the
scenario may be a bit more palatable if we do not have to
wave away a super-Eddington accretion rate in this way.  Or put
another way, it is simply interesting to note that the scenario
can work with a sub-Eddington time-averaged accretion rate as
long as equations~(\ref{eqn:dtr}) and (\ref{eqn:Eddrequire}) are
satisfied, or as long as
\begin{equation}
     \zeta \gtrsim \frac{1.9\times 10^{-5}}{\tau_4 M_3}.
\label{eqn:zeta}
\end{equation}
This quantity must be $\zeta \leq 1$, and is estimated
to be $\zeta\sim 0.1$ \citep{Farrar:2014yla} (although that is a
value for the average over all SMBHs, and does not necessarily
apply to a single SMBH).  Such a value is easily accommodated if
$\tau_4\sim 1$, as we might expect for UHECR protons, and even
fits for heavier nuclei, for which $\tau_4\sim 2.5\times 10^{-3}$.

We have thus shown that the TA hot spot can be explained as
a roughly steady-state phenomenon by the sub-Eddington capture
and tidal disruption of stars by the SMBH in M82.  The scenario
works independent of whether the UHECRs are protons or iron
nuclei, although the timescale parameter space is a bit narrower
for iron nuclei, a consequence of the larger deflection of iron
nuclei in the Milky Way magnetic field.

\section{Isotropic flux}
\label{sec:isotropic}

We now investigate whether the isotropic UHECR flux implied by
this scenario is consistent with that observed under the
assumption that the UHECR luminosity of M82 (and of Cen A) are
fairly typical for such SMBHs.  This analysis applies not only
to the hypothesis that TDEs are responsible for the hot and warm
spots, but to any scenario in which there are hot/warm spots
associated with Cen A and M82.

We begin with a simple analysis.  The isotropic-equivalent
luminosities of M82 and Cen A are, respectively, $2.9\times
10^{43}\, {\rm GeV}\, {\rm s}^{-1}$ and 
$1.4\times 10^{43}\, {\rm GeV}\, {\rm s}^{-1}$.  Both SMBHs
are at a distance $R\lesssim 4$ Mpc, and so the UHECR luminosity
density in a 4-Mpc sphere around us is $\rho_L \simeq 5.4 \times
10^{-33} {\rm GeV}\, {\rm cm}^{-2}\, {\rm s}^{-1}$.  If the
UHECR emissions from Cen A and M82 are both beamed into a
fraction $\Omega_{\rm jet}$ of the $4\pi$ solid angle, then
$\rho_L$ is reduced by $\Omega_{\rm jet}$.  If M82 and Cen A are
not atypical, though, then there must be $\sim \Omega_{\rm
jet}^{-1}$ other beamed UHECR sources, aimed in other
directions, for every source that we see.  This then cancels the
$\Omega_{\rm jet}$ beaming reduction leaving $\rho_L$
unchanged.  
Since both Cen A and M82 appear, in the jetted-TDE scenario, to
be aimed at us, we infer that $\Omega_{\rm jet}$ is unlikely to
be small in this scenario.  The tension we will find below
between the hot/warm-spot fluxes and the isotropic intensity can
be relaxed, though, if both Cen A and M82 just happen to be
highly beamed and both in our direction.
If our local neighborhood is not atypical, then $\rho_L$
provides an estimate of the universal UHECR luminosity density.
If the local density is greater by a factor $f_\rho$ than the
cosmic mean density, then the universal UHECR luminosity density
is $\rho_L/f_\rho$.

The isotropic UHECR intensity (energy per unit area per unit
time per unit solid angle) is
\begin{equation}
     I = \int_0^R\, dr\, r^2\, f(r) \frac{\rho_L}{4 \pi r^2} =
     \frac{\rho_L}{4\pi} \int_0^R\, dr\, f(r) = \frac{\rho_L
     R}{8\pi},
\label{eqn:Iintegral}
\end{equation}
where $R$ is the GZK radius, and the second equality is obtained
by approximating the fraction of UHECR energy
emitted at a distance $r$ that makes it to us to be  $f(r)\simeq
1-(r/R)$ \citep{Kotera:2011cp}.  If the TA hot spot and PAO
warm spot are real and attributed to M82 and Cen A,
respectively, then the isotropic UHECR flux should be $I = 1.37
\times 10^{-7}\,F_{1.7}\,f_\rho^{-1}\,{\rm GeV}\, {\rm cm}^{-2}\, {\rm
s}^{-1}\, {\rm sr}^{-1}$.  This is, for $f_\rho=1$, 16 times
greater than the isotropic intensity $I_{o} = 7.9 \times
\ten{-9}$ GeV cm\(^{-2}\) s\(^{-1}\) sr\(^{-1}\).  The
discrepancy cannot be alleviated with a smaller value of
$F_{1.7}$ because, as discussed after
equation~(\ref{eq:hotSpotObs}), we are using the specific
intensities at $E \simeq 10^{19.5}$\, EeV, which are fairly well
determined, as proxies for the full energy flux and isotropic
intensity.  

It is, however, likely that the tension can be
alleviated, at least in part, with a value $f_\rho>1$.  The
local density is uncertain, but as one indication of the value
of $f_\rho$, we can use the total SMBH in the $R\simeq$Mpc
sphere, assuming that the UHECR luminosity density is
proportional to the density of mass SMBHs.  In addition to the
SMBHs in Cen A and M82, there is
also the $\sim4\times10^6\,M_\odot$ SMBH in the Milky Way and
the $\sim10^8\,M_\odot$ SMBH in Andromeda, as well as a
$\sim10^6\,M_\odot$ SMBH in M32.  This totals to
$\sim2\times10^8\,M_\odot$ in SMBHs within a distance $R\simeq4$
Mpc implying a local SMBH density $\simeq 7.5\times 10^5\,
M_\odot$~Mpc$^{-3}$, roughly 3 times the universal SMBH density
$\simeq 2.9\times 10^5\, M_\odot$~Mpc$^{-3}$
\citep{Dzanovic:2006px}.  There is still residual factor of
$\sim5$ discrepancy that remains, even accounting for this
$f_\rho\sim 3$, that must be accounted for if the TDE explanation
for the TA and PAO hot spots is to remain viable.  This level of
discrepancy is we believe, given the order-of-magnitude nature
of the analysis, as well as the measurement and astrophysical
uncertainties, not necessarily fatal for the TDE scenario.
The local luminosity density $\rho_L$ we inferred could have
been reduced a bit by considering a sphere of slightly larger
radius; there are uncertainties almost of order unity in the
measured fluxes; and the Poisson fluctuation in our inference of
$\rho_L$ is also of order unity.

So far we have been using the UHECR flux from M82 and Cen A to
infer a luminosity density, and the uncertainty from
small-number statistics has been noted above.  There is,
however, an additional uncertainty that may arise from the
dependence of the mean TDE rate on SMBH mass.  SMBHs are
distributed with a mass function $dn/dM$
\citep{Dzanovic:2006px,Caramete:2009nh}, and there is evidence
that the TDE rate varies with the SMBH mass.  We infer an UHECR
luminosity density from measurement of the UHECR flux from one
or two $\sim3\times 10^7$ SMBHs.  Suppose, though, that the TDE
rate varies as $\Gamma(M) =\Gamma(M=3\times 10^7\,M_\odot)
(M/3\times 10^7\,M_\odot)^{-\beta}$.  The luminosity density we
infer from the measured M82 flux would then be $L_{\rm tde} \int
(dn/dM) (M/3\times 10^7\,M_\odot)^{-\beta}$, where $L_{\rm tde}$
is the UHECR luminosity from one burst.  If we then use the
best estimate $\beta \simeq 0.22$ from \citet{Stone:2014wxa},
the  SMBH mass function from \citet{Dzanovic:2006px}, and
integrate from $10^5\,M_\odot$ (below which there is little
evidence for SMBHs) to $10^8\,M_\odot$ (above which stars will
be swallowed without being tidally disrupted
\citep{Magorrian:1999vm}), we find---unfortunately for the
TDE scenario---a luminosity density $\sim1.7$ times higher.
This power-law index $\beta$ is, however, quite uncertain, and
if we suppose that it is instead $\beta\simeq0.5$, then the
inferred luminosity density is decreased by $\sim0.5$.  This may
thus provide some wiggle room for the tension between the M82
and Cen A fluxes and the isotropic intensity, although is
unlikely to be the entire explanation.  Changes to the upper and
lower limits of integration do not alter this conclusion.  We do
note that the masses of the SMBHs in Cen A and M82 are quite
similar, both around $(3-5)\times10^7\,M_\odot$.  If, for some
reason, the TDE rate were to be maximized for SMBHs of this mass,
and smaller for SMBHs of both lower and higher masses, then
the universal UHECR luminosity could be reduced significantly
relative to what we inferred above.  In this case, the high
fluxes toward M82 and Cen, relative to the isotropic intensity,
would be a consequence of our chance proximity to two SMBHs of
this specific mass.

The tension between the hot/warm-spot fluxes and the isotropic
intensity may also be relaxed if UHECR consist at the source, at
least in part, of other nuclei, like helium, carbon, nitrogen, or
oxygen.  The path length of such nuclei through the
intergalactic medium is far smaller than the $\sim 200$ Mpc
GZK distance of protons and iron nuclei \citep{Kotera:2011cp}.
If there is significant UHECR production in such nuclei, then
the isotropic intensity inferred from the measured
$D\lesssim4$~Mpc luminosity density will be smaller.  Such a
scenario implies a different observed UHECR composition in the
hot/warm spots and in the isotropic component.  There may
already be some evidence for intermediate-mass nuclei in UHECRs
\cite{Aab:2014aea}.

\section{Discussion: TDE scorecard} 
\label{sec:discuss}

Here we have investigated the possibility that tidal disruption
events fueled by the accretion of stars onto the SMBH in M82
could account for the hot spot reported by the Telescope Array
and that TDEs onto the SMBH in Cen A could explain the warm spot
seen by the Pierre Auger Observatory toward Cen A.  Given the
measurement uncertainties and considerable astrophysical
uncertainties, it is difficult to make precise statements about
the viability of the scenario.  Although there are some tensions
at the order-of-magnitude level, there is, as far as we can
tell, no silver bullet that rules the scenario out at the level
of more than an order of magnitude.

Our conclusions are as follows:  Energetics make it unlikely,
although not impossible, that the hot spot towared M82 is the
result of a single burst, a tension that is probably greater if
UHECRs are protons rather than iron nuclei.  Dispersion in
galactic and intergalactic magnetic fields disperse the UHECR
arrival times.  This magnetic-dispersion time is, if anything,
likely higher for protons than for iron nuclei.  The
single-burst scenario is also unappealing as it implies that the
hot spot is evanescent, something that we see as a chance
occurrence.  This chance event is made even less likely if the
warm spot toward Cen A is also explained another chance event.

The energetics requirements are relaxed, though, if the UHECRs
in the hot spot result from a number of TDEs in M82 that have
occurred over a magnetic dispersion time, a scenario in which
the UHECR fluxes in the hot/warm spots are roughly in steady
state.  The required efficiency of UHECR production in each TDE
event can then be reduced at the expense of an increased TDE
rate.  We do show, though, that the TDE rates can still remain
low enough so that the time-averaged accretion rate in M82
remains sub-Eddington, something that may be desirable, though
not necessarily required, to explain the quiescent nature of the
SMBH in M82.  (This is less of a concern, of course, for Cen A,
which is quite active.)  This latter, softer, requirement, is
satisfied, though, only at the expense of introducing a slight
tension in the required UHECR efficiency per TDE.  That tension
can be reduced if the TDE is highly beamed.  Significant beaming
introduces, however, the notion that the UHECR flux from M82
results from our chance position within the TDE's jet, an
ingredient that is less appealing if we must also explain the
PAO warm spot in terms of TDEs from Cen A's SMBH.  Any
significant beaming requirement for Cen A would also be more
difficult given that the radio observed jet in Cen A is not
pointed toward us.

We note that the time between jetted TDEs in our scenario is a
bit higher than the rate expected from existing TDE statistics.
Scalings between TDE rates and SMBH masses derived in 
\citet{Stone:2014wxa} suggest that the characteristic time
between TDEs in a $3\times 10^7\,M_\odot$ SMBH is
$\Gamma^{-1}\sim10^4$ yr.  \citet{Farrar:2014yla} estimate
further that only a 
fraction $\zeta\sim 0.1$ of TDEs are jetted.  If we take this
value for M82, then the time between UHECR-producing events is
greater than the magnetic-dispersion time.  There are, however,
considerable uncertainties in these estimates, and there may
also be considerable variation between the jetted fraction for
one particular SMBH and the mean inferred by averaging over all
SMBHs.  

We then investigated the isotropic flux of UHECRs that is
expected if the sources of UHECRs in M82 and Cen A are not
atypical.  This analysis applies not only to the hypothesis that
the UHECR sources in M82 and Cen A are TDEs, but to any scenario
in which there are hot/warm spots from Cen A and M82.  The
observed UHECR fluxes from M82 and Cen 
A imply a local UHECR luminosity density.  We find that
if the universal UHECR luminosity density is taken to be this
local luminosity density, then the isotropic UHECR intensity is
about 16 times larger than that observed.  There is, however,
some evidence that the local mass density in SMBHs is higher,
perhaps by $\sim3$, than the universal density.  Even so, there
is still a tension, at the $\sim5$ level, between the
hot/warm spot fluxes and the isotropic intensity.  Possible
explanations for this residual tension may arise from our
underestimate of the local overdensity; small-number statistics
in the number of SMBHs; uncertainties in the characterization
of the hot/warm spots; a mixed composition of
UHECRs including intermediate-mass nuclei with smaller
GZK cutoffs; and/or some SMBH-mass dependence of the
TDE rate.

It is interesting to wonder whether the SMBH $\sim 4 \times
10^6\,M_\odot$ SMBH at the center of Milky Way
\citep{Ghez:2008ms} should produce UHECRs.  The answer is
probably not.  Assuming the Milky Way is a core galaxy, the
expected time, from \citet{Stone:2014wxa}, between TDEs for the
Milky Way's SMBH is $3.9 \times10^4$ yr.  As discussed above,
the magnetic-dispersion time within the Milky Way can be, for
reasonable magnetic-field parameters, quite a bit smaller than
this.  It is thus not surprising that we do not see an UHECR hot
spot toward the Galactic center, even if our SMBH does produce
TDEs at the expected rate.

Future measurements should help shed additional light on the
viability of TDEs as the sources of UHECRs.  The viability of
the TDE scenario for the isotropic flux has been discussed in
\citet{Farrar:2008ex,Farrar:2014yla}, but if the hot/warm spots
are real and attributed to M82 and Cen A, then there are
additional challenges discussed above.  It will be interesting
to see if the evidence for the hot and warm spots continues with
additional data, and if so, the characterization of those fluxes
should improve.  There may be differences, which we will explore
elsewhere, in the energy distribution of
UHECRs in the hot/warm spots, that come from 3.8 Mpc, versus
those in the rest of the sky, which come from much greater
distances and thus experience greater photo-pion absorption.

There may also be signatures in ultra-high-energy neutrinos
expected to be produced alongside UHECRs.  Although the the
arrival times of UHECRs are spread out by magnetic fields,
neutrinos travel in a straight path.  While some UHE neutrinos
may be produced during photo-pion or photo-disintegration in
transit from the source, neutrinos produced during the TDE
should arrive in a single burst, roughly coincident with the TDE
light arrival time.  Detection of UHE neutrinos from a TDE (any
TDE, not necessarily one in M82 or Cen A) would thus
significantly bolster the case for TDEs as a source of
UHECRs.  However, given the low fluxes and TDE rates,
nonobservation of neutrinos from TDEs is unlikely to constrain
the scenario.

Before closing, we speculate on the possibility that the IMBH in
M82 \cite{Patruno:2006bw,Pasham:2015tca} (should the evidence for that IMBH
survive) may have something to do with the TA hot spot.  It may
be possible for IMBHs to produce their own TDEs.  A TDE would
need to occur close to the innermost stable circular orbit of
the IMBH in order to accelerate the in-falling matter and
produce a flare.  A main-sequence star would be disrupted before
then, but a white dwarf could possibly survive until it gets
close enough to an IMBH to produce a TDE with an intense flare.
Another possibility is that IMBHs might perturb the orbits of
stars in a way similar to the Kozai mechanism
\citep{Perets:2006bz}, and thus increase the rate of TDEs in the
host galaxy.  The tension with the isotropic UHECR intensity
might thus be explained by an IMBH-enhanced TDE rate in M82
relative to what it would be otherwise.

\section*{Acknowledgments} \label{sec:ack}

The authors would like to thank Julian Krolik, Joe Silk, Meng Su
and Ilias Cholis for useful discussions.  This work was
supported by NSF Grant No. 0244990, NASA NNX15AB18G, the John
Templeton Foundation, and the Simons Foundation.

\end{document}